\documentclass{PoS}

\usepackage{amsmath}
\usepackage{graphicx}
\usepackage{graphics}
\usepackage{subfig}

\title{Measuring luminosity at LHCb using elastic diphoton dimuon production}

\ShortTitle{Measuring luminosity at LHCb using elastic diphoton dimuon production}

\author{\speaker{Dermot Moran (on behalf of the LHCb collaboration}\thanks{The author would like to acknowledge the support of Science Foundation Ireland}\\
        University College Dublin\\
        E-mail: \email{dermot.moran@ucdconnect.ie}}

\abstract{We report on an indirect method being used to measure luminosity at LHCb. It involves recording the event rate of elastic diphoton dimuon production. Preliminary MC studies suggest that with 1 fb$^{-1}$ of data this method could provide a luminosity measurement with a precision of better than 2\%.}

\FullConference{XVIII International Workshop on Deep-Inelastic Scattering and Related Subjects, DIS 2010\\
		April 19-23, 2010\\
		Firenze, Italy}

\begin{document}

\section{Introduction}

LHCb is one of the experiments located on the Large Hadron Collider at CERN. Its primary goal is to investigate rare and CP violating B hadron decays. At LHC energies $b\bar{b} $ pairs are predominantly produced at small polar angles, i.e either in the forward or backward region. The geometry of the LHCb detector exploits this production topology. It is a forward arm spectrometer covering a pseudorapidity of $1.9 < \eta  < 4.9$. Collisions with a centre of mass energy of 7 TeV have begun at the LHC. It is expected that 1 fb$^{-1}$ of data will be collected over the next 2 years.

A precise measurement of the integrated luminosity is required to make cross-section measurements. A direct luminosity measurement involves measuring the beam current and shape. The beam shape can be determined using the Van Der Meer Scan method \cite{VanDerMeer}, or from the reconstruction of beam gas interactions using the Vertex Locator (VELO) \cite{BeamGas}. Using the VELO in this way the integrated luminosity for 2009 has been measured to be 6.8 +- 1 ub$^{-1}$. The uncertainty of 15\% is dominated by measurement of the beam currents. This is expected to improve in 2010 as the beam current measurement system was designed for higher energy beams than were used in 2009. The beams of 2009 were also broader than nominal. For this reason the VELO was not fully closed around the interaction region resulting in a worse than nominal precision. The VELO has been fully closed for the current 7TeV run. Based on these improved running conditions, it is expected that an integrated luminosity measurement for 2010 will be made with a precision of 5\%.

An indirect method for determining the luminosity involves measuring the number of events, N, seen for a particular process with a precisely known cross-section, $\sigma$. The luminosity, L, is then determined by the equation $L = N/\sigma$. A suitable process for this method is elastic diphoton dimuon production as it has a relatively high cross-section which is known with a theoretical uncertainty of 0.4\%. In the following we outline our triggering and selection strategies for these events at LHCb. A full study of the signal and its backgrounds has been conducted at a centre of mass energy of 10 TeV and we have scaled the results of this study to 7 TeV. 



\section{Triggering on diphoton dimuon events at LHCb}

The LHCb trigger consists of a hardware trigger, Level 0 (L0), and 2 software triggers, High Level Trigger 1(HLT1) and High Level Trigger 2 (HLT2). To select diphoton dimuon events the L0 trigger looks for low multiplicity events containing 2 muon objects (aligned sequence of hits in the muon chambers) with Pt > 80 MeV. The multiplicity of an event is determined using the Sintillator Pad Detector (SPD) which is part of the LHCb calorimeter system. Charged particles crossing the SPD produce ionization, from the resulting scintillation light the multiplicity of an event can be determined. 

The HLT1 trigger looks to fully reconstruct a dimuon using the L0 muon objects and information from the VELO and the tracking stations. If the reconstructed dimuon has a mass greater than 1 GeV the next level of the software trigger, HLT2, is run.
HLT2 applies a full event reconstruction. If an event contains exactly two muons with a dimuon Pt  less than 900 MeV and a distance of closest approach less than 150 um than it is written to disk. Based on LPAIR \cite{LPAIR} predictions and a full LHCb detector simulation we expect diphoton dimuon events to be triggered at LHCb with an effective cross-section of 77 pb. 

\section{Backgrounds}

 The backgrounds considered are inelastic diphoton dimuon production, dimuons produced from double pomeron exchange (DPE), dimuons from exclusive $J/\psi$ and $\psi\prime$ decay, dimuons from $J/\psi$ decay, dimuons from the Drell-Yan process, dimuons produced from heavy quark pair decay ($b\bar{b} $ and $c\bar{c} $) where both quarks decay semi-leptonically to muons, and pion kaon mis-id.

The contribution from inelastic diphoton dimuon production has been investigated using LPAIR. In these events one (semi-inelastic) or both (fully-inelastic) of the protons dissociate. The uncertainty on the cross-section for these two processes has been estimated using a number of different proton structure functions. The semi-inelastic cross-section has been estimated to have an uncertainty of 25\% while the fully-inelastic cross-section has been estimated to have an uncertainty of 50\%. 

Dimuons resulting from DPE have been studied using the Cox-Forshaw model of POMWIG \cite{POMWIG} and the Boonekamp Peschanski Royon (BPR) model of DPEMC \cite{DPEMC}. An uncertainty on the DPE cross-section of 50\% has been evaluated using both models and a number of different pomeron PDFs. 

Exclusive $J/\psi$ and $\psi\prime$ photoproduction (photon pomeron fusion) have been evaluated at 14 TeV using the STARLIGHT generator \cite{STARLIGHT}. These events are interesting in their own right as they may provide evidence for the existence of the odderon (C-odd partner of the pomeron). It has been observed that suitable mass cuts around the $J/\psi$ and $\psi\prime$ mass regions completely remove these backgrounds. It is assumed that this cut will have the same effect at 7 TeV.

All of the above have been studied using a full LHCb detector simulation. The backgrounds from $J/\psi$ decay,  the Drell-Yan process, heavy quark ($b\bar{b} $,$c\bar{c} $) semi-leptonic decays and pion kaon mis-id have been evaluated at 4 vector level using pythia. For these processes we have assumed reconstruction and triggering efficiencies of 100\%. When evaluating the pion kaon mis-id rate we considered the main sources of mis-id to be punch through and decay in flight.We estimate that the pion kaon mis-id rate will be determined from data with an uncertainty of 10\%.

\begin{figure}
 \centering
 \subfloat[Charged particle multiplicity distribution]{\label{fig:tiger}\includegraphics[width=0.525\textwidth]{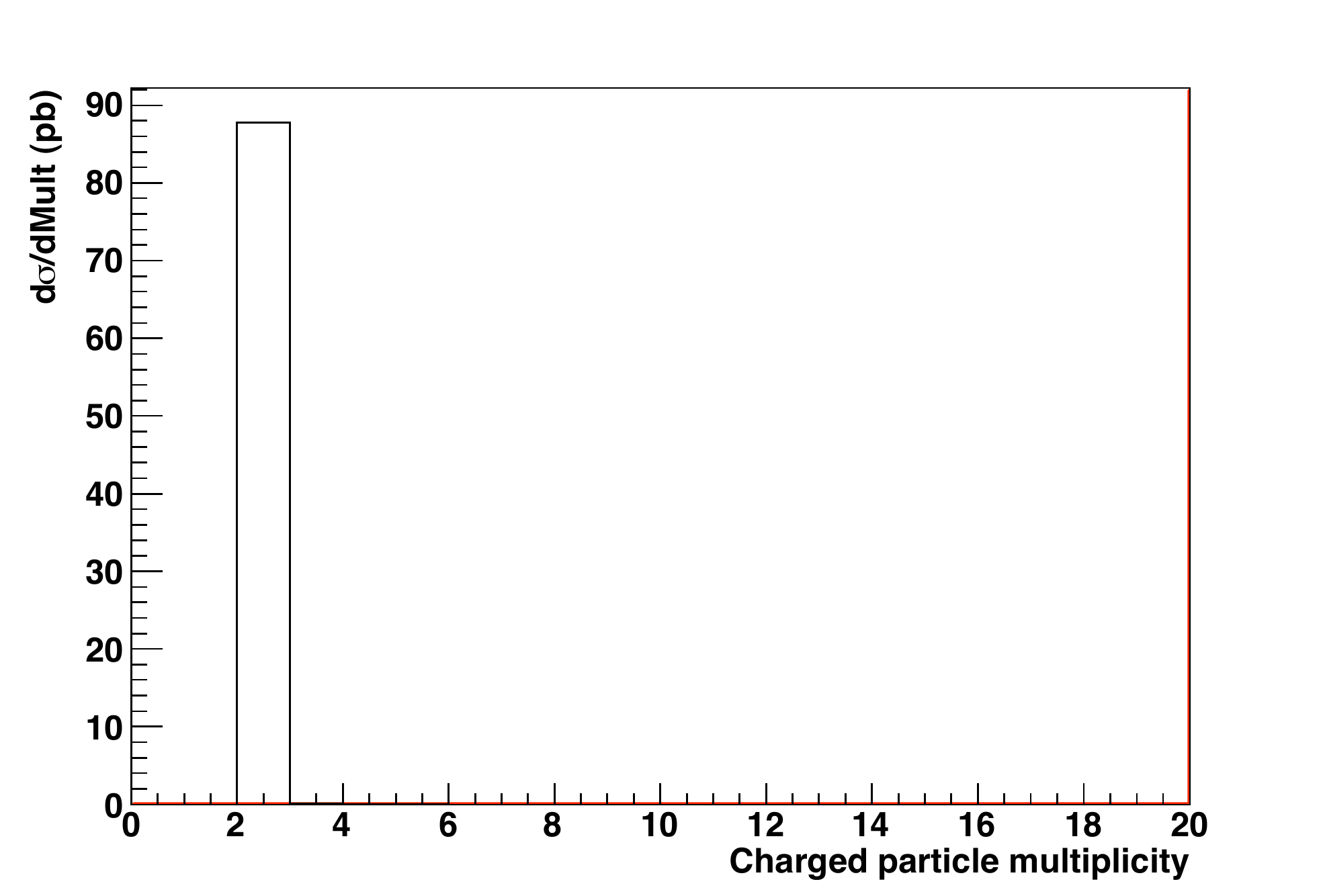}}                
 \subfloat[Dimuon transverse momentum distribution]{\label{fig:tiger}\includegraphics[width=0.525\textwidth]{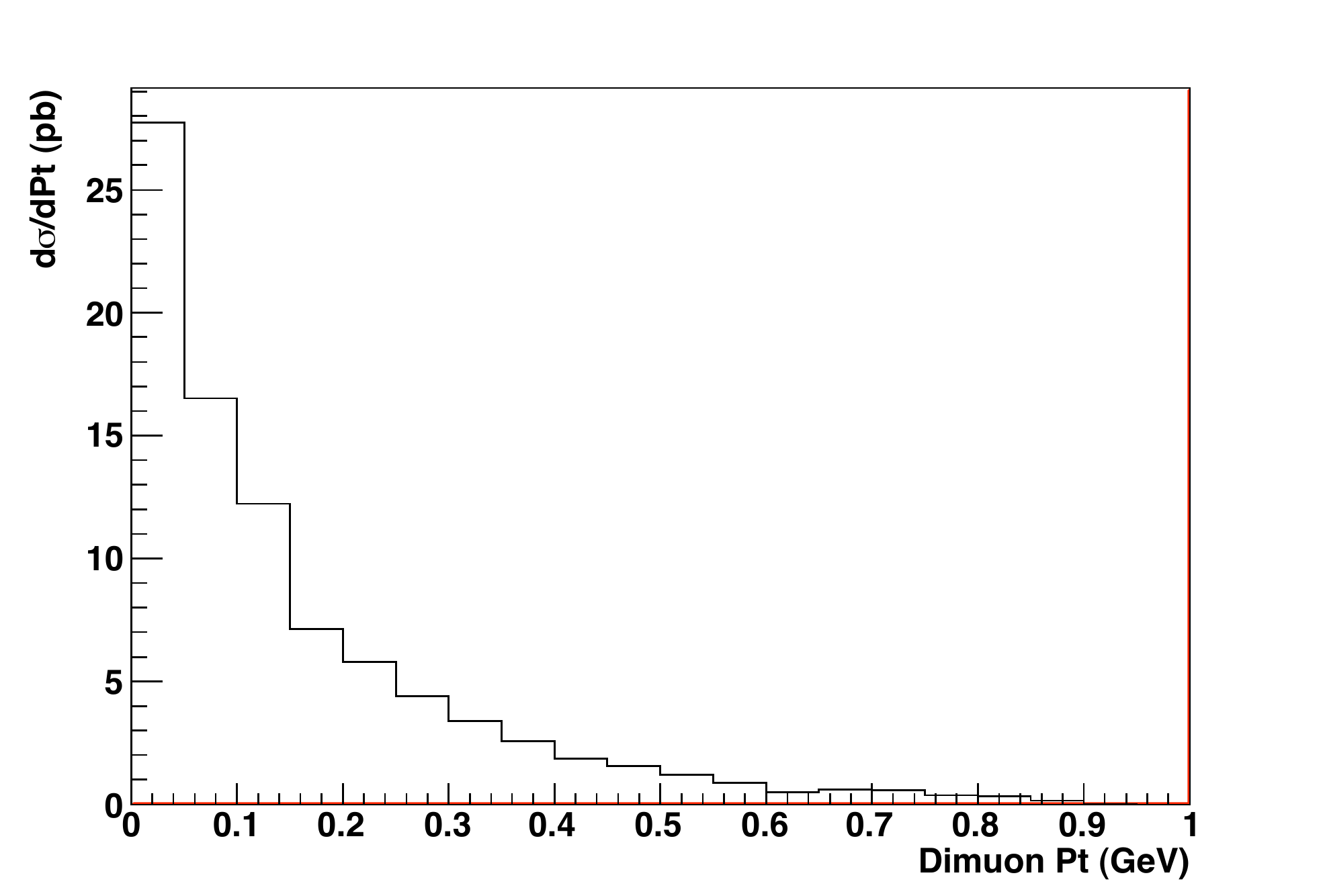}}
 \caption{ Diphoton dimuon distributions for offline reconstructed events at LHCb}
 \label{fig:Wselscheme}
\end{figure}

\section{Offline Selection}

For the purpose of isolating the signal diphoton dimuon events from the various backgrounds an offline selection has been developed which exploits some of the signals unique features. As the protons do not dissociate the events contain only two muons and therefore have a relatively small charged particle multiplicity (see Figure 1(a)). The muons are generally produced back to back in the transverse plane and therefore the dimuon has a comparatively small Pt (see Figure 1(b)).

 The following cuts are used in our offline selection:

\vspace{5pt}
1) The dimuon transverse momentum must satisfy Pt < 50 MeV.

2) The muon transverse momentum must satisfy Pt > 1 GeV.

3) There must be exactly 2 reconstructed charged particles in the event.

4) The dimuon Mass must be between 2.6 GeV and 20 GeV.

5) The dimuon mass regions 3 - 3.2 GeV and 3.6 - 3.8 GeV are omitted.
\vspace{5pt}

Table 1 shows the expected number of events in 1 fb$^{-1}$ of data after this offline selection is applied. The statistical uncertainties and the systematic uncertainties due to the cross-section predictions are also shown. Using this offline selection the signal events can be selected with a purity of 94 \%. Assuming detector systematics of 0.8\% we expect we can make  a luminosity measurement with a precision of 1.8 \%. This uncertainty is dominated by the systematic uncertainty on the predicted DPE cross-section and by our knowledge of the muon trigger and reconstruction efficiencies. We expect recent measurements of exclusive processes at CDF will lead to an improvement in the precision of the DPE cross-section prediction.

\begin{table}
\caption{Expected number of events for 1fb$^{-1}$ of data running at 7TeV}  
\centering       

\begin{tabular}{l  rrrr}
\\ [0.2ex]
\hline\hline                         
 Process  &       Events in 1 fb$^{-1}$  &      Stastical uncertainty   &     Systematic uncertainty &\\ [0.55ex]
\hline\hline                     

Diphoton dimuon & 5512 & +/-74 & +/-22 & \\[-1ex]
\\ [0.2ex]

 \hline\hline  

DPE dimuon & 57  & +/-8 & +/-29 &\\[-1ex]
\\ [0.2ex]
\hline

 Semi-inelastic dimuon & 24  & +/-5 & +/-6 &\\[-1ex]
\\ [0.2ex]
\hline

Fully-inelastic dimuon & 1  & +/-1 & +/-1 &\\[-1ex]
\\ [0.2ex]
\hline
Drell-Yan dimuon & 17  & +/-4 & +/-1 &\\[-1ex]
\\ [0.2ex]
\hline
$J/\psi$ dimuon & 0  & +/-1 & +/-1 &\\[-1ex]
  \\ [0.2ex]
  \hline
$c\bar{c}$ dimuon & 0  & +/-1 & +/-1 &\\[-1ex]
  \\ [0.2ex]
  \hline
 $b\bar{b}$ dimuon & 0  & +/-1 & +/-1 &\\[-1ex]
 \\ [0.2ex]
 \hline
Pion/Kaon mis-id & 270  & +/-16 & +/-27 &\\[-1ex]
\\ [0.2ex]
\hline\hline  
Total Background & 369  & +/-19 & +/-40 &\\[-1ex]
  \\ [0.2ex]

\hline\hline                             
\end{tabular}
\label{tab:PPer}
\end{table}

\section{Conclusion}

The integrated luminosity for 2009 has been measured with a precision of 15\% using VELO reconstructed beam gas interactions. It is expected that with the improved running conditions of 2010 the integrated luminosity for the current run will be measured with a precision of 5\%.  
Preliminary MC studies show that with 1 fb$^{-1}$ of data an integrated luminosity measurement with a precision of less than 2\% can be made by recording the event rate of diphoton dimuon production. This uncertainty is dominated by the systematic uncertainty on the predicted DPE cross-section and by our knowledge of the muon trigger and reconstruction efficiencies. It is expected that 1 fb$^{-1}$ of data will be collected over the next 12-18 months.

\end{document}